\documentclass[referee]{aa}
\usepackage{graphicx}
\usepackage{txfonts}
\begin{document}

   \title{Multi-wavelength analysis of the field of the dark burst GRB 031220}


\author{A. Melandri \inst{1,2}, B. Gendre \inst{3},
            L. A. Antonelli \inst{1}, A. Grazian \inst{1}, A. de
            Ugarte Postigo \inst{4}, J. Gorosabel \inst{4}, L. Piro
            \inst{3}, G. Kosugi \inst{5}, N. Kawai \inst{6}, M. de
            Pasquale \inst{3}, G. P. Garmire \inst{7} }

\offprints{A. Melandri \\ e-mail: melandri@mporzio.astro.it}

\institute{INAF - Osservatorio Astronomico di Roma, Via Frascati 33,
              00040, Monte Porzio Catone - Italy\\
              \email{melandri@mporzio.astro.it} \and {Dipartimento di
              Fisica, Universit\`a di Cagliari, S. P. Monserrato,
              09042 Cagliari - Italy} \and {INAF - Istituto di
              Astrofisica Spaziale e Fisica Cosmica, Via Fosso del
              Cavaliere 100, 00133, Roma - Italy } \and {Istituto de
              Astrofisica de Andalucia (IAA-CSIC), Camino Bajo de
              Huetor 24, 18.008 Granada - Spain} \and {Subaru
              Telescope - National Astronomical Observatory of Japan,
              Hilo, HI 96720 - USA} \and {Department of Physics -
              Tokyo Institute of Technology, Meguro, Tokyo 152-0033}
              \and {Department of Astronomy and Astrophysics, 525
              Davey Lab, Pennsylvania State University, University
              Park, PA 16802-6305, USA} }

\date{}

\abstract{
  
  We have collected and analyzed data taken in different spectral
  bands (from X-ray to optical and infrared) of the field of GRB
  031220 and we present results of such multiband
  observations. Comparison between images taken at different epochs in
  the same filters did not reveal any strong variable source in the
  field of this burst. X-ray analysis shows that only two of the seven
  Chandra sources have a significant flux decrease and seem to be
  the most likely afterglow candidates. Both sources do not show the
  typical values of the $R - K'$ colour but they appear to be
  redder. However, only one source has an X-ray decay index (1.3 $\pm$
  0.1) that is typical for observed afterglows. We assume that this
  source is the best afterglow candidate and we estimate a redshift of
  1.90 $\pm$ 0.30. Photometric analysis and redshift estimation for
  this object suggest that this GRB can be classified as a {\it Dark
  Burst} and that the obscuration is the result of dust extinction in
  the circum burst medium or inside the host galaxy.  }

\authorrunning{A. Melandri {\it et al.}  }
\titlerunning{Multiwavelength analysis of the field of GRB 031220 : a
dark burst}
          
\maketitle

%

\section{Introduction}

The lack of optical afterglow for a large fraction (about 50\%) of
well localized X-ray afterglows and, sometimes, together with the
detection of radio afterglow, leads to the definition of the
phenomenological class of the so called {\it Dark Bursts}. The nature
of this class of events, bursts with no apparent optical afterglow, is
still not clear but recent works have suggested some possible
scenarios. In some cases the non-detection of the optical transient
could be simply due to the lack of suitable instrumentation: the
slowness of the response or the depth of the surveying combined with
some dim or rapid decaying event could bias the determination of the
truly {\it Dark Bursts} population (Berger {\it et al.} 2002, Fynbo
{\it et al.}  2001). It is also possible that this kind of bursts have
an intrinsically fainter optical afterglow compared to the afterglow
of other GRBs at all wavelengths (De Pasquale {\it et al.} 2003,
Jakobsson {\it et al.} 2004) and this could happen if the afterglow
decelerated in a low density medium (Sari {\it et al.} 1998). If
instead GRBs are associated with the death of massive stars
(Paczy\'nski 1998, MacFadyen \& Woosley 1999, Wheeler {\it et al.}
2000) then the optical flux of the afterglow could be blocked by a
large fraction of interstellar dust along the line of sight (Stratta
{\it et al.} 2004) or the single event could be the result of a highly
absorbed star formation burst placed in a dusty molecular cloud
(Reichart \& Price 2002, Lazzati {\it et al.}  2002).  From X-ray
analysis (Galama \& Wijers 2001) there is evidence for high column
densities of gas close to GRBs, but the measured optical extinction is
smaller than expected because the hard $\gamma$-ray radiation of the
burst destroys the dust in their enviroment. In this {\it obscured}
scenario, the failed detection of the optical transient could be
easily ascribed to extinction by the dust of the host galaxy.
Moreover, is now well assessed that GRBs are the most energetic events
in the universe and they have a large redshift distribution ($0.1 < z
< 4.5$) (e.g., Schaefer, Deng \& Band 2001; Bagoly {\it et al.}
2003). So a fraction of {\it Dark Bursts} could be originated at
high-z ($z > 5$, Lamb \& Reichart 2000, Wijers {\it et al.} 1998) and
their emission could be dumped by the Lyman absorption redshifted to
the optical-infrared bands. Obviously, also a combination of these
effects can determine the {\it dark} nature of these events. However,
for a few events (potentially {\it Dark Bursts}) of which the distance
has been determinated, the contribution of high redshift effects to
the optical darkness of this events is very little or negligible and
they are probably the result of dust extinction in the circum burst
medium (Djorgovski {\it et al.}  2001, Piro {\it et al.} 2002).


\section{GRB 031220}

GRB 031220 was detected by
HETE2\footnote[1]{http://space.mit.edu/HETE/Bursts/GRB031220/}
satellite on 2003 December 20.1458 UT (HETE2 trigger 2976) in the
6-120 keV band with a  count rate of 466 counts $s^{-1}$.
 The center coordinates of the Soft X-ray Camera (SXC) error
circle are RA: $+04^{h} 39^{m} 34.3^{s}$, DEC: $+07^{\circ} 22^{'}
25^{''}$ (J2000). The peak energy ($E_{peak}$) for this burst was
49.24 keV with a fluence of $1.946 \times 10^{-6} erg~cm^{-2}$ and
lasted-- 23.7 seconds (HETE2 trigger 2976).

The first observation in the optical band of the field of this GRB was
performed by the ROTSE-IIIb telescope $\sim1.9$ hours after the burst
and did not reveal any new source up to an unfiltered magnitude of
19.2 (Rykoff {\it et al.}, 2004). For comparison the magnitude of GRB
030329 afterglow faded to a magnitude $\sim 12.5$ in the optical band
$\sim1.5$ hours after the burst (Price \& Peterson 2003, Smith {\it et
al.} 2003). The HETE2 error circle (20.77 arcmin of diameter) was
imaged with many different optical telescopes since early times until
some months after the burst (Kosugi {\it et al.}  2003, Antonelli {\it
et al.} 2003, Gorosabel {\it et al.} 2004) but no optical afterglow
has ever been identified. The SXC error box was completely covered
also by the Chandra X-ray Observatory with two set of observations.


\section{Follow up observation and data analysis}

\subsection{{\bf X-ray data}}

   \begin{table*}
      \caption[]{Right Ascension and Declination (J2000) of the
Chandra sources detected inside the HETE2 error circle as reported by
De Pasquale {\it et al.} 2003, GCN 2502. In last two column we have
reported the count rate for all detected sources in the first and
second Chandra observation. The reported upper limit for the non detection 
of source \# 1 in the second observation is quoted at the 95 \% confidence level.}
         \label{chandra}
\begin{center}
\begin{tabular}{c c c c c c}
            \hline \hline $\#$ & Source name & Right Ascension &
            Declination & First Observation & Second Observation\\ & &
            & & Count rate (10$^{-3}$count s$^{-1}$)& Count rate
            (10$^{-3}$count s$^{-1}$)\\ \hline 1 & CXOU
            J043944.3+072036 & $+04^{h} 39^{m} 44.35^{s}$ &
            $+07^{\circ} 20^{'} 36.74^{''}$ & 0.8 $\pm$ 0.2 & $<0.2$\\ 2
            & CXOU J043939.7+072318 & $+04^{h} 39^{m} 39.77^{s}$ &
            $+07^{\circ} 23^{'} 18.64^{''}$ & 1.3 $\pm$ 0.2 & 0.8
            $\pm$ 0.3 \\ 6 & CXOU J043946.4+072220 & $+04^{h} 39^{m}
            46.47^{s}$ & $+07^{\circ} 22^{'} 20.19^{''}$ & 2.6 $\pm$
            0.3 & 2.5 $\pm$ 0.4 \\ 7 & CXOU J043946.1+072256 &
            $+04^{h} 39^{m} 46.14^{s}$ & $+07^{\circ} 22^{'}
            56.46^{''}$ & 0.8 $\pm$ 0.2 & 0.4 $\pm$ 0.2 \\ 17 & CXOU
            J043954.8+072149 & $+04^{h} 39^{m} 54.83^{s}$ &
            $+07^{\circ} 21^{'} 49.33^{''}$ & 0.7 $\pm$ 0.2 & 0.4
            $\pm$ 0.3 \\ 27 & CXOU J044003.7+072055 & $+04^{h} 40^{m}
            03.76^{s}$ & $+07^{\circ} 20^{'} 55.26^{''}$ & (2.1 $\pm$
            0.3) & Extended object\\ 37 & CXOU J043857.8+072449 &
            $+04^{h} 38^{m} 57.14^{s}$ & $+07^{\circ} 24^{'}
            48.84^{''}$ & Border of the chip& Spurious detection \\
            \hline 55 & CXOU J043929.2+072314.& $+04^{h} 39^{m}
            29.20^{s}$ & $+07^{\circ} 23^{'} 14.88^{''}$ & 0.5 $\pm$
            0.2 & --- \\ 58 & CXOU J044012.5+071945 & $+04^{h} 40^{m}
            12.58^{s}$ & $+07^{\circ} 19^{'} 45.20^{''}$ & 0.6 $\pm$
            0.2 & --- \\ 66 & CXOU J043927.2+072351 & $+04^{h} 39^{m}
            27.23^{s}$ & $+07^{\circ} 23^{'} 51.04^{''}$ & 0.3 $\pm$
            0.1 & --- \\ 78 & CXOU J043934.4+072124 & $+04^{h} 39^{m}
            34.49^{s}$ & $+07^{\circ} 21^{'} 24.16^{''}$ & 0.3 $\pm$
            0.2 & --- \\ 82 & CXOU J043855.2+072457 & $+04^{h} 38^{m}
            55.23^{s}$ & $+07^{\circ} 24^{'} 57.79^{''}$ & 0.3 $\pm$
            0.1 & --- \\ \hline \hline
         \end{tabular}
\end{center}
     \end{table*}

X-ray observations were carried out with the Chandra space observatory
in two different epochs: 5.62 (De Pasquale {\it et al.}, 2003) and
28.47 (Gendre {\it et al.}, 2004) days after the burst. Observations
were performed using ACIS\_I detectors and exposure times were 40 and
20 kiloseconds respectively. The data reduction was performed using
the version 3.0 of the CIAO software. All the events flagged as bad by
the calibration chain have been discarded and only the grades 0, 2, 3,
4 and 6 (within the provided good time intervals according to the CIAO
cookbook\footnote[2]{see the Chandra Proposer Observatory Guide
available from the Chandra web site for details.})  have been
kept. The event file has been checked for flaring background activity
and no such event has been found. Then, the event file has been
filtered for energies between 0.2 and 8.0 keV and this filtered event
file has been used for detection with the wavelet tool {\it
wavdetect}.

During the first Chandra observation, eleven sources were detected
inside the HETE2 error box plus one source (source \#~37) near the
border of the chip. Sources position extracted from De Pasquale {\it
et al.} (2003) and Gendre {\it et al.}  (2004) are reported in Table
\ref{chandra}. Five of these sources (sources \#~55, \#~58, \#~66,
\#~78, and \#~82) were too faint to be detectable during the second
Chandra observation. Source \#~37, reported by De Pasquale {\it et
al.}  (2003), is not detected in our refined analysis (Gendre {\it et
al.}  2004). This source was spurious, its false detection was mainly
due to the vicinity of a CCD edge. On the remaining six sources, five
were detected. The only exception is source \#~1, which disapeared. We
indicate in Table \ref{chandra} the observed count rate of each source
during the first and second observation. The unabsorbed flux in units
of erg s$^{-1}$ cm$^{-2}$ can be obtained by multiplying the values
indicated in Table \ref{chandra} by $1.47 \times 10^{-11}$ (0.2-8.0
keV band) or $6.40 \times 10^{-12}$ (2.0-10.0 keV band). To derive
these conversion factors, we used a power law model with a photon
index of 2.

\begin{figure}
   \centering \includegraphics[width=9cm]{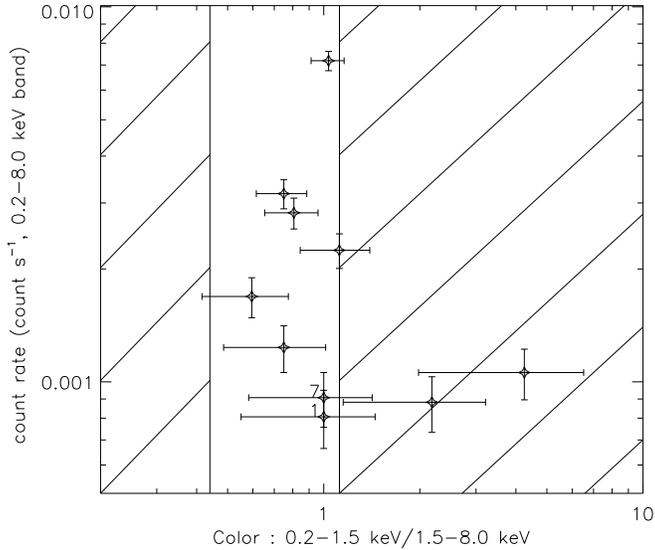} \caption{X-ray
   colour of the variable Chandra sources within the field of view. We
   have indicated the name of variable sources \#~1 and \#~7 with a
   number. The two vertical lines correspond to a power law with a
   photon index of 2 (left line) and 3 (right line). The expected
   colour value for a typical afterglow is between these two
   lines. Both candidates have a colour value within the expected
   range.  } \label{fig_x_1} \end{figure}

Due to the poor signal of possible afterglow candidates we could not
extract source spectra in order to analyze them (e.g. Ballet
2003). Thus we used soft (0.2-1.5 keV) and hard (1.5-8.0 keV) X-ray
bands to derive candidates colours. Results for the first Chandra
observation are plotted in Fig. \ref{fig_x_1} together with the valid
colour range expected from typical afterglow. All candidates are
compatible with a power law with photon index $\sim 2-3$ and a
Galactic column density of hydrogen atoms $N_{H}$ of $\sim 1.14 \times
10^{21}$ (Dickey \& Lockman 1990).

\begin{figure}
   \centering \includegraphics[width=9cm]{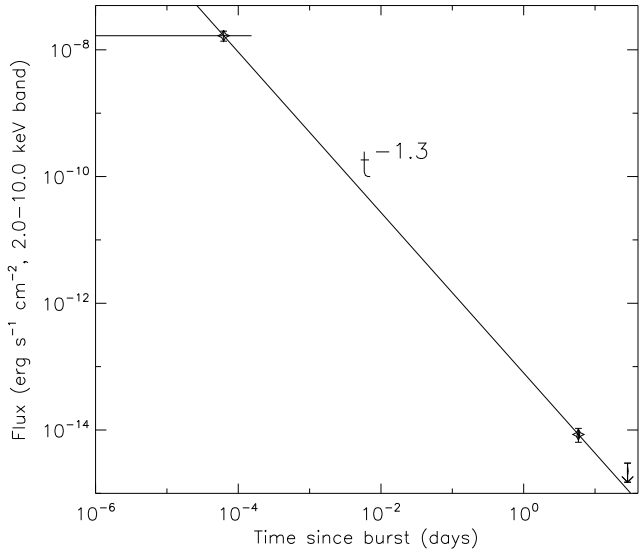}
\caption{Light curve of the source \#~1 in the 2.0-10.0 keV band,
together with the best fitted power law model ($F_{x} \propto
t^{-1.3}$). We have indicated the HETE2(+FREGATE) prompt flux and the
two Chandra observations. Note that this source is not detected in the
second Chandra observation.  } \label{fig_x_2} \end{figure}

We used the data from the two Chandra observations to look for 
variability. Among the six sources detected in the first Chandra observation
and bright enough to be detected in the second observation only 2
of them show a flux decrease by more than a factor of 2 
(at a 2 $\sigma$ level of significance) : source \#~1 is not detected and source
\#~7 is marginally detected. The corresponding flux variation factor is $>$ 2.6
and 2.3 for sources \#1 and \#7 respectively (Gendre {\it et al.}, 2004). The remaining
sources are constant (within error bars). Note that source \#~27
appears to be extended in the second Chandra observation, and thus its
count rate is not totaly reliable. It is compatible with the one
measured in the first Chandra observation.
We then built light curves (in the 2.0-10.0 keV band), using the value of the
prompt emission, recorded by the French Gamma Telescope (FREGATE) on board HETE2 
(the flux was $1.7 \times 10^{-8} \pm 0.3 \times 10^{-8}$ in the 2.0-10.0
keV band, Atteia, private communication), and each Chandra observation 
as a single bin in the light curve.  In Fig. \ref{fig_x_2} we
show the light curve of the afterglow candidate source \#~1 together
with the best fitted power law relationship (with a decay index of
$1.3 \pm 0.1$).  The light curve of the afterglow candidate source \#
7 is flatter, with a decay index of  $1.20 \pm 0.05$. The source \#7 decay 
index within the two chandra observations is $0.7_{-0.5}^{+0.1}$.

\subsection{{\bf Optical data}}

   \begin{table} 
    \begin{minipage}[t]{\columnwidth} 
    \caption{Observation log for GRB 031220 field. Observing dates are
   listed chronologically starting from December 2003 to March 2004
   and are referred to the beginning of the exposures. $^{*}$)
   Effective wavelength for the four BUSCA channels are 3654.5 (ch a),
   5007.4 (ch b), 6455.8 (ch c) and 7999.7 (ch d) \AA, corresponding
   to the U, B, R and I band ~respectively.}

    \label{journal}
    \begin{tabular}{c c c c c c}
    \hline \hline \bf Date & $\Delta$ T & \bf Filter & \bf Exp & \bf
    Seeing & \bf Obs\\ (UT) & (days) & & (s) & ($^{\prime \prime}$) &
    \\ \hline Dec 20.4722 & 0.3264 & $I$ & 5x600 & 0.9 &
    \footnote{Subaru Telescope + FOCAS}\\ Dec 21.4636 & 1.3088 & $I$ &
    5x600 & 0.9 & $^{a}$\\ Dec 28.0375 & 7.8917 & $R$ & 3600 & 1.1 &
    \footnote{TNG Telescope + DOLORES}\\ Dec 30.9636 & 10.8178 & $K'$
    & 3600 & 1.7 & \footnote{TNG Telescope + NICS} \\ \hline Jan
    04.8695 & 15.7237 & $H$ & 4980 & 0.9 & \footnote{CA3.5m Telescope
    + OMEGA-Prime}\\ Jan 05.8773 & 16.7315 & $J$ & 4500 & 1.0 &
    $^{d}$\\ Jan 06.9488 & 17.8030 & $K'$ & 4440 & 0.9 & $^{d}$\\
    \hline Mar 07.7858 & 78.6400 & $J$ & 4080 & 1.0 & $^{d}$\\ Mar
    08.8796 & 79.7338 & $H$ & 2700 & 1.1 & $^{d}$\\ Mar 20.8082 &
    91.6622 & ch a,b,c,d$^{*}$ & 3600 & 1.4 & \footnote{CA2.2m
    Telescope + BUSCA}\\ Mar 21.8016 & 92.6558 & ch a,b,c,d & 3600 &
    1.4 & $^{e}$\\ Mar 22.8034 & 93.6576 & ch a,b,c,d & 3600 & 2.5 &
    $^{e}$\\ Mar 23.8067 & 94.6609 & ch a,b,c,d & 3600 & 2.1 &
    $^{e}$\\ \hline
    \end{tabular}
  \end{minipage}
\end{table}

Optical and infrared observations of GRB 031220 field are listed in
Table \ref{journal}. Different epoch images, in different spectral
bands, taken with different telescopes and instrumentation (as
reported in Table \ref{journal}) have been collected and analyzed.

\subsubsection{{\it Subaru observations}}

The Subaru observation consists of two data sets of I-band images
taken at two different epochs, approximately 8 and 32 hours after the
burst. Images have been acquired with the FOCAS instrument operating
in imaging mode on the Subaru 8.2m telescope on Mauna Kea
(Hawaii). The two sets of images start at 11:20 UT on Dec 20 (Kosugi
{\it et al.} 2004) and 11:07 UT on Dec 21 respectively. The field was
covered with 5 pointings of 600 seconds for each data set obtained
moving the center position of each exposure in order to cover most of
the SXC error circle of GRB 031220 (20.77 arcmin of diameter). The
field of view for a single image is $\sim$ 6 arcmin but all the
Chandra sources listed in Table \ref{chandra} are present in each set
of images.

\subsubsection{{\it TNG observations}}

Two observations were taken with the Telescopio Nazionale Galileo
(TNG) at La Palma (Canary Islands, Spain): a first exposure (3600
seconds) in the R band started at 00:54 UT on 28 Dec with the DOLORES
instrument operating in imaging mode (Antonelli {\it et al.} 2003) and
a second exposure (3600 seconds) in the K' band started at 23:07 UT on
30 Dec with the NICS instrument.

\subsubsection{{\it Calar Alto observations}}

Three sets of observations were taken at different epochs at the Calar
Alto Observatory in Spain. First and second sets were taken on the
first days of January (4, 5 and 6) and March (6 and 7) respectively
with the OMEGA-prime wide-field near infrared camera (JHK' bands on
January and JH bands on March) at the Calar Alto 3.5m telescope. A
further set of images were taken at the end of March with the BUSCA
camera ($\sim$ 12 arcmin square of field of view) that is a CCD system
which allows simultaneous direct imaging of the same sky area in four
colours, corresponding to the U, B, R and I band. The effective
wavelengths for BUSCA filters are reported in the caption of Table
\ref{journal}.

\subsection{Image photometry}

  \begin{figure}
   \centering
   \includegraphics[width=9cm]{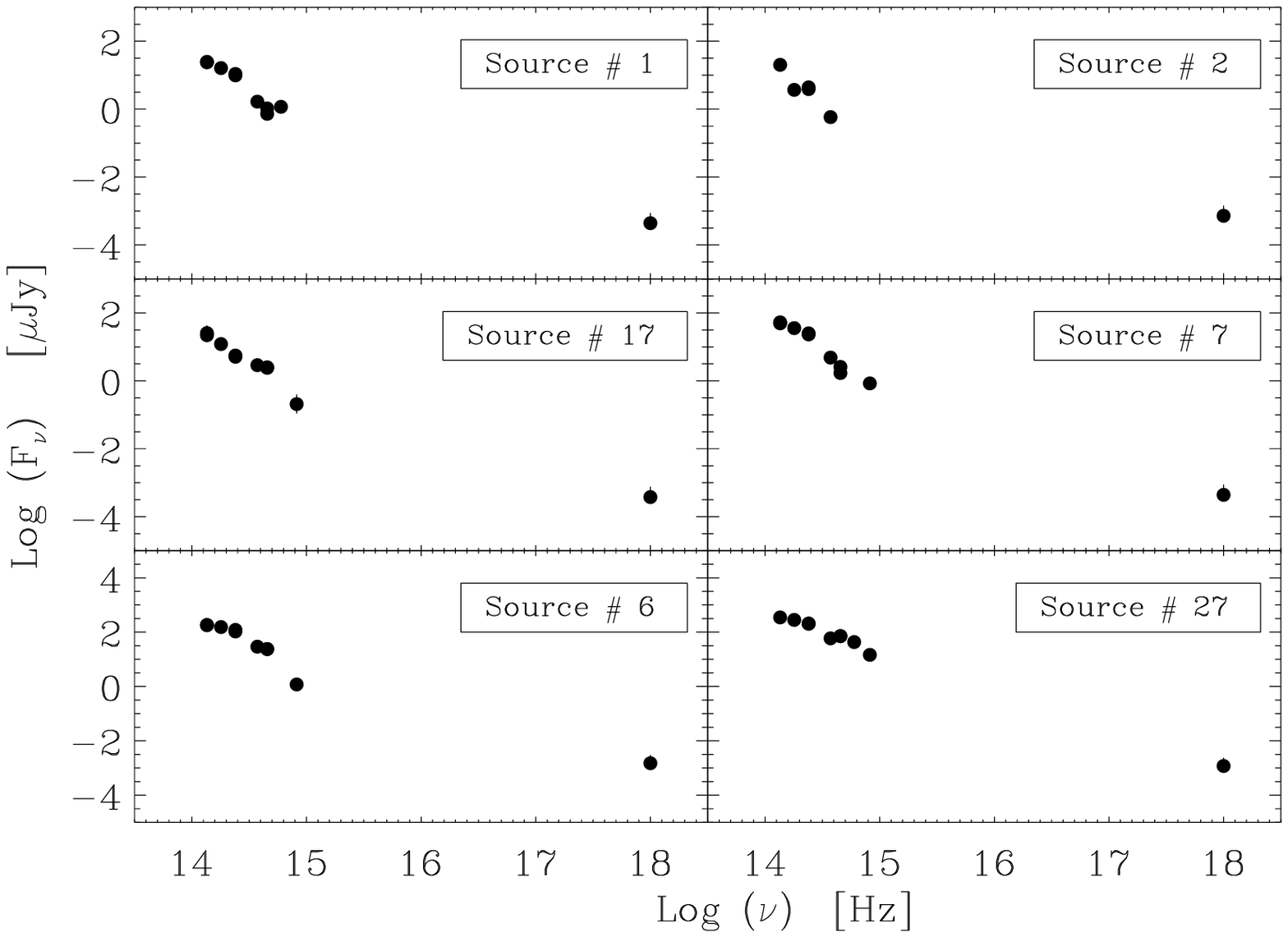}
      \caption{Spectral energy distribution for the six brightest
      Chandra sources analyzed. Infrared (K'HJ), optical (IRBV) and
      X-ray fluxes of the first Chandra observation are plotted with filled circles.}
         \label{flux}
   \end{figure}

The Chandra source \#~37 is far from other candidates, it is not
present in all images and the refined analysis of the X-ray data after
the second Chandra observation excluded it as a possible variable
source in the field (Gendre {\it et al.}, 2004). For this reason our
analysis has been limited to the other six Chandra sources, for which
we show in Fig. \ref{flux} the derived spectral energy distribution.

The first step of our analysis was to investigate of the field of GRB
031220 in order to check the possible presence of new variable
sources, namely a possible optical afterglow without X-ray
emission. In order to do this, we make comparison between the two
Subaru observations (taken at $\sim8$ and $\sim32$ hours after the
burst) and then between Subaru observations and R band observation
from TNG (taken $\sim8$ days after the burst) or BUSCA I band
observation (taken $\sim3$ months after the burst). First of all, a
visual comparison between two exposures of the same portion of the
field has been made looking for any possible transient object. Then,
for each image a list of objects has been extracted using Sextractor
(Bertin \& Arnouts 1996) and compared each other in order to
investigate all the objects present in the field. This deeper
inspection did not reveal variable sources at a significance level of
2 $\sigma$.

Then the photometric analysis of the six brightest Chandra sources
listed in Table \ref{chandra} has been performed. Optical fields have
been calibrated using USNO catalogues (as BUSCA filters are not
standard we have assumed the correspondence $ch~a = U$, $ch~b = B$,
$ch~c = R$, $ch~d = I$) whereas for infrared fields 2MASS catalog has
been used.  The photometric analysis of each single image has been
performed in the IRAF \footnote[3]{IRAF is the Image Reduction and
Analysis Facility distributed by the National Optical Astronomy
Bservatories (NOAO), which are operated by AURA Inc., under
cooperative agreement with US National Science Foundation.} enviroment
using the DAOPHOT package (Stetson 1987). The magnitudes for all
afterglow candidates, listed in Table \ref{tabmag}, have been
estimated making differential aperture photometry with the PHOT
routine. In particular, the magnitudes of a set of selected and not
saturated stars have been estimated, compared with known catalog
values for these stars and finally used to evaluate the magnitudes of
all the candidates. Using this procedure we can use small apertures 
for each candidate in order to avoid contaminations in the measured 
magnitudes and colours from near objects. All the magnitudes calculated 
and listed in Table \ref{tabmag} have been corrected in each filter for 
the corresponding Galactic extinction, $A_{\lambda}$, reported in the 
same table ($E_{(B-V)}$ = 0.146, Schlegel {\it et al.} 1998).

\begin{table*}
  \begin{center}
    \caption{Optical and infrared magnitudes of the six afetrglow
    candidates corrected for galactic extinction. In the last two
    lines the colours $R - K'$ and $J - K'$ are reported, calculated
    using the two TNG observations and one CA3.5m observation, that
    are best accurate measurements in these bands. In the last two
    columns the limiting magnitude $M_{lim}$ (at 5$\sigma$) and the
    magnitude system correction $M_{corr}$ for different filters are
    reported.}
    \label{tabmag}
    \begin{tabular}{c c c c c c c c c c}
    \hline
    \hline

\bf Source & \#~1 & \#~2 & \#~6 & \#~7 & \#~17 & \#~27 & $A_{\lambda}$
& $M_{lim}$ & $M_{corr}$\\ \hline \bf Filter (Obs) & & Mag & $\pm$ &
Err & & & &\\ \hline \hline

$U$ (e) & --- & --- & 22.95 $\pm$ 0.12 & 23.33 $\pm$ 0.17 & --- &
20.23 $\pm$ 0.02 & 0.72 & 25.2 & -0.533\\

$B$ (e) & 23.77 $\pm$ 0.36 & --- & --- & --- & --- & 19.85 $\pm$ 0.10
& 0.63 & 24.7 & +0.085\\

$R$ (b) & 23.61 $\pm$ 0.11 & --- & 20.22 $\pm$ 0.05 & 23.09 $\pm$ 0.11
& 22.71 $\pm$ 0.07 & 19.03 $\pm$ 0.05 & 0.38 & 24.0 & -0.209\\

$R$ (e) & 23.43 $\pm$ 0.19 & 24.20 $\pm$ 0.33 & 20.19 $\pm$ 0.12 &
22.34 $\pm$ 0.13 & --- & 19.18 $\pm$ 0.11 & 0.38 & 24.8 & -0.143\\

$I$ (a) & 23.00 $\pm$ 0.26 & 24.13 $\pm$ 0.26 & 19.89 $\pm$ 0.26 &
21.83 $\pm$ 0.26 & 22.40 $\pm$ 0.25 & 19.12 $\pm$ 0.25 & 0.29 & 26.5 &
-0.448\\

$I$ (a) & 23.43 $\pm$ 0.26 & 24.23 $\pm$ 0.27 & 19.93 $\pm$ 0.25 &
21.81 $\pm$ 0.25 & 22.46 $\pm$ 0.25 & 19.11 $\pm$ 0.25 & 0.29 & 26.5 &
-0.448\\

$I$ (e) & 23.04 $\pm$ 0.21 & --- & 20.19 $\pm$ 0.20 & 22.06 $\pm$ 0.21
& --- & 19.37 $\pm$ 0.20 & 0.29 & 23.5 & -0.448\\

$J$ (d) & 20.52 $\pm$ 0.09 & 21.40 $\pm$ 0.42 & 17.94 $\pm$ 0.04 &
19.59 $\pm$ 0.05 & 21.23 $\pm$ 0.15 & 17.22 $\pm$ 0.23 & 0.13 & 22.0 &
-0.896\\

$J$ (d) & 20.41 $\pm$ 0.26 & 21.53 $\pm$ 0.42 & 17.80 $\pm$ 0.23 &
19.52 $\pm$ 0.24 & 21.15 $\pm$ 0.34 & 17.22 $\pm$ 0.20 & 0.13 & 21.7 &
-0.896\\

$H$ (d) & 19.55 $\pm$ 0.14 & 21.15 $\pm$ 0.34 & 17.11 $\pm$ 0.12 &
18.69 $\pm$ 0.12 & 19.87 $\pm$ 0.14 & 16.45 $\pm$ 0.12 & 0.08 & 21.4 &
-1.360\\

$K'$ (c) & 18.57 $\pm$ 0.14 & --- & 16.41 $\pm$ 0.08 & 17.75 $\pm$
0.10 & 18.70 $\pm$ 0.19 & --- & 0.08 & 20.9 & -1.846\\

$K'$ (d) & 18.61 $\pm$ 0.27 & 18.79 $\pm$ 0.27 & 16.40 $\pm$ 0.25 &
17.82 $\pm$ 0.26 & 18.54 $\pm$ 0.56 & 15.70 $\pm$ 0.25 & 0.05 & 20.6 &
-1.846\\

    \hline \hline

$R - K'$ & 5.04 $\pm$ 0.17 & --- & 3.81 $\pm$ 0.09 & 5.34 $\pm$ 0.14 &
4.01 $\pm$ 0.20 & --- \\ $J - K'$ & 1.95 $\pm$ 0.16 & --- & 1.53 $\pm$
0.09 & 1.84 $\pm$ 0.11 & 2.53 $\pm$ 0.24 & --- \\

    \hline \hline
    \end{tabular}
   \end{center}
  \end{table*}

\section{Redshift estimation}

Photometric redshifts have been estimated for all the Chandra sources
by adopting a $\chi^{2}$ minimization technique of the observed
Spectral Energy Distribution (SED) on a spectral library drawn from
the Rocca-Volmerange synthesis models (Le Borgne \& Rocca-Volmerange
2002, Fioc \& Rocca-Volmerange 1997) as described by Fontana {\it et
al.} (2000). This method takes into account the star formation history
of each galaxy type, the reddening produced by internal dust and Lyman
absorption produced by intergalactic dust. This is a widely used and
well tested technique for redshift determination (Fernandez-Soto {\it
et al.} 1999, Csabai {\it et al.}  2000, Rowan-Robinson 2003).

We further tested this method with a GRB event that is at high
redshift, applying this tecnique to GRB 000131, the burst with the
highest known redshift ($z = 4.5$, Andersen {\it et al.}  2000). We
used the result of broad band photometry extrapolated by the author
and we obtained a redshift of $4.65 \pm 0.20$ for this event, in good
agreement with Andersen's one.

\begin{figure}[!ht]
   \centering \includegraphics[width=9.1cm,height=14.5cm]{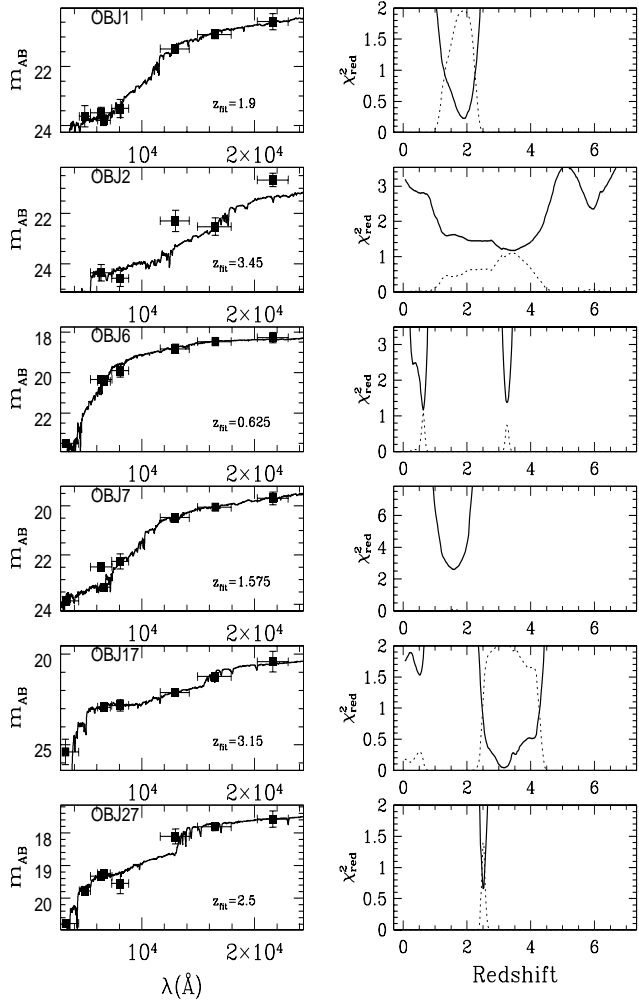}
      \caption{Magnitudes best fit (left side) and $\chi^{2}_{red}$
      distribution versus redshift (right side) for the six Chandra
      candidates. Filled squares represent the observed spectral
      energy distribution (horizontal bars show the amplitude of the
      filter while vertical bars indicate the errors in magnitude). In
      the right panel, solid lines are the best fit model that
      minimizes the $\chi^{2}_{red}$, whereas dashed lines represent
      the probability distribution not normalized. }
         \label{red}
   \end{figure}

Do to the lack of a clearly fading afterglow we can assume that our
estimated magnitudes reported in Tab.\ref{tabmag} have to be ascribed
to the host galaxy of GRB 031220. Because of the photometric redshift
determination algorithm requires input magnitudes in the AB system we
converted the measured magnitudes from the Vega system to AB
system. We used the relation $M_{AB}=M_{Vega}-M_{corr}$, where
$M_{corr}$ are the corrections applied and listed in Table
\ref{tabmag}. The results of redshifts determination procedure are
summarized in Table \ref{redvalue} and shown in Fig. \ref{red}: left
plot of the figure shows the AB magnitudes of the six Chandra
candidates in different photometric bands with the corresponding best
fit curve; in the right plot is visible the $\chi^{2}$ reduced
distribution versus the photometric redshift $Z_{phot}$. As reported
in Table \ref{redvalue}, for our two potential afterglow candidates,
source \#~1 and source \#~7, we estimated a redshift of $1.90 \pm
0.30$ ($\chi^{2}_{red}$ = 0.22 for 7 degrees of freedom) and $1.57 \pm
0.45$ ($\chi^{2}_{red}$ = 2.60 for 7 degrees of freedom)
respectively. In the same table we have reported the minimum of the
reduced $\chi^{2}$ distribution, the probability $P_{\chi^{2}_{red}}$
and the rest-frame colour excess $E_{(B-V)}$ for each candidate. The
estimate of the colour excess with this algorithm is obtained assuming
a Small Magellanic Cloud (SMC) extinction law.

To be sure that we can really exclude high redshifts, we apply the fit
procedure on this two candidates forcing the redshift $Z_{phot}$ to be
$\ge 2.5$. After the test, for these two objects we obtain a
$\chi^{2}$ reduced distribution very broad: in particular we found
$Z_{phot} = 2.50$ for source \#~1 and $Z_{phot} = 5.25$ for source \#
7 and the correspondent values of $\chi^{2}_{red}$ were 2.79 and 13.60
respectively (for 7 degrees of freedom).

\begin{table}
 \begin{center}
     \caption{Photometric redshift estimation ($Z_{phot}$) for the six
candidates, with the correspondent minimum of the reduced $\chi^{2}$
distribution, the probability $P_{\chi^{2}_{red}}$ and the rest-frame
colour excess $E_{(B-V)}$.}
    \label{redvalue}
    \begin{tabular}{c c c c c}
    \hline \hline

\bf Candidate & $\bf Z_{phot}$ & $\bf \chi^{2}_{red}$ & $\bf
P_{\chi^{2}_{red}}$ & $\bf E_{(B-V)}$ \\ \hline $\#$ 1 & 1.90 $\pm$
0.30 & 0.22 & 0.98 & 0.10 \\ $\#$ 2 & 3.45 $\pm$ 0.90 & 1.17 & 0.31 &
0.00 \\ $\#$ 6 & 0.62 $\pm$ 0.15 & 1.16 & 0.32 & 0.85 \\ $\#$ 7 & 1.57
$\pm$ 0.45 & 2.60 & 0.01 & 0.03 \\ $\#$ 17 & 3.15 $\pm$ 0.55 & 0.04 &
0.99 & 0.00 \\ $\#$ 27 & 2.50 $\pm$ 0.10 & 0.66 & 0.70 & 0.00\\

    \hline \hline
    \end{tabular}
   \end{center}
  \end{table}

\section{Discussion}

We analyzed the field of GRB 031220. We find seven X-ray sources
inside the HETE2 error circle and only two of them (source \#~1 and
source \#~7) are showing a fading behaviour. 

Taking into account the prompt and the two Chandra observations,
source \#~1 has a X-ray decay index of $1.3 \pm 0.1$, that is typical for observed
afterglows at early time (De Pasquale et al. 2005), and source \#7 got a decay index 
of $1.2 \pm 0.05$. One can note that the afterglow light curve cannot be always 
extrapolated backward up to the prompt emission, which can lie above or below the 
extrapolated light curve (see e.g. Costa et al. 1997) Thus, taking only into account 
the two Chandra observations, source \#1 has a decay index of at least 1.1 (90\% 
confidence level) and source \#7 a shallow decay of $0.7_{-0.5}^{+0.1}$ (90\% confidence 
level). After 5 days, the observed usual decay is $\sim$~1.9 (Gendre et al. 2005), 
clearly not consistent with the observed shallow decay of source \#7. This shallow 
decay is also not consistent with the decay inferred from the optical variation 
($0.21 \pm 0.05$, Gorosabel {\it et al.}, 2004). It may be due to some flaring activities, 
like \object{GRB 970508} (Piro et al. 1998), but most of the afterglows present only a 
smooth power law decay with sometime a possible jet break (De Pasquale et al. 2005, Gendre 
et al. 2005). From all of these considerations, we propose that source \#1 is the afterglow 
of \object{GRB 031220} rather than source \#7, while we cannot formally exclude that source 
\#7 is related to \object{GRB 031220}.

The estimation of $Z_{phot}$ for source \#~1 ($1.90 \pm 0.30$) is
compatible with the value of $2.3 \pm 0.5$ found using the Boer \&
Gendre relationship (Gendre \& Boer 2005) and also with the pseudo
redshift of $\sim 1.95$ estimated from the prompt emission (Atteia
2003). It should be note that the Boer \& Gendre redshift estimator is
based on the flux estimation and the error quoted include the error
due to the uncertainty on the flux calculation. Nevertheless source
\#~ 1 and source \#~7 got a similar flux so the redshift estimator is
very similar for the 2 sources.

The more significant afterglow candidate found by the X-ray data
analysis (source \#~1) has peculiar value of $R - K'$ colour, that is
$\sim 5$. This value is still compatible with the optical transients
colour-colour selection criteria of Gorosabel {\it et al.} (2002) but
it is not typical. This potential disagreement with the colours
expected for GRB afterglows (Gorosabel {\it et al.} 2002) is likely
because the $R$ and $K'$-band fluxes of object \#~1 is (if not
totally) dominated by a constant host galaxy component. Thus, our 
estimated colour should be considerd like an upper limit for the afterglow 
and most properly referred to the host galaxy of this GRB. Is interesting
to note that the $R - K'$ colour of source \#~1 is redder that the
host galaxy sample by Le Floc'h {\it et al.} (2003), explainable by a
high dust content in this GRB host galaxy candidate. From the
literature the mean of $R - K'$ colour distribution for optically
obscured burst is $\sim 3$ (Le Floc'h {\it et al.} 2003). The $R - K'$
colour found for our source indicates that this burst should be a high
obscured or a high redshift event. As an example, the first well
localized burst with no optical afterglow (GRB 970828, $z = 0.9578$),
have $R - K' = 3.7$ and this value is well explained by dust
extinction within the host galaxy (Djorgovski {\it et al.} 2001). Also
the dark GRB 000210 (Piro {\it et al.} 2002) shows $R - K' = 2.5$ and
the darkness is explained by the effect of local absorber within the
host galaxy (Gorosabel {\it et al.} 2003).  Nevertherless, the burst
with the highest spectroscopic redshift ($z = 4.5$, GRB 000131,
Andersen {\it et al.}  2000) has a similar value of $R - K'$ colour of
$\sim 3.7$. Then, with the only $R - K'$ colour information it is
impossible to discriminate between the possibility of high obscured or
high redshift event. However, the results obtained forcing the fit
procedure with high values of redshift, described at the end of
previous section, permit us to exclude high redshift for this event.

De Pasquale {\it et al.} (2003) shows that the X-ray flux of optically
dark bursts is on average weaker than the flux of bright bursts. They
also find that about 20\% of the dark bursts show an optical to X-ray
flux much lower than that observed in optically bright events,
corresponding to a $\beta_{ox} < 0.6$, where $\beta_{ox}$ is defined
as the ratio between the optical and X-ray spectral index. Moreover,
in a recent work, Jakobsson {\it et al.} (2004) shows that optically
dark bursts display the trend to be located in a well defined area on
a plane $F_{opt}$ versus $F_{X}$, where $F_{opt}$ and $F_{X}$ are the
optical (in the R band) and the X-ray flux respectively. This area is
located below the constant line $\beta_{ox} = 0.5$ and this value
describe the transition between dark and luminous optical burst
(Jakobsson {\it et al.} 2004).  For GRB 031220 we do not see an
optical fading afterglow but only the host galaxy and our measured
magnitudes should be considered like upper limits for this burst.
Using the value of the magnitude measured in the R (or I) band and the
X-ray flux found in our analysis for source \#~1 we can extropolated
X-ray and optical flux at 11 hours after the burst assuming a
decay index of $\alpha = 1$.. We obtain an X-ray flux of $Log F_{X} 
\sim -1.50$ ( identical value for source \#~7 ) whereas 
the upper limit for the optical flux is $Log F_{opt} \sim 0.01$. 
If instead we assume the magnitude of source \#~7 we find an upper 
limit for the optical flux of $Log F_{opt} \sim 0.1$. As one can see 
in Fig.1 of Jakobsson {\it et al.} (2004) source \#~1 is located close 
to the transition area defined by $\beta_{ox}$, inside the belongings 
region for the {\it Dark Bursts},  whereas source \#~7 is outside 
this region. This information, together with the fact that we did not
reveal any optical afterglow emission, allows us to classify this
burst as a {\it Dark} event.

On the basis of the colour excess found in our analysis we can
estimate the rest-frame column density of hydrogen atoms
($N_{H}^{~z}$) for source \#~1. Assuming a SMC-like interstellar
medium ($N_{H}= 4.9~ \times 10^{22}~ E_{(B-V)}~ cm^{-2}$) we obtained 
 $N_{H}^{~z} ~(~\#~1~) = (0.5 \pm 0.2)~ \times 10^{22}~ cm^{-2}$ 
for source \#~1.  Instead, for source \#~7 we obtained 
$N_{H}^{~z} ~(~\#~7~) = (0.1 \pm 0.2)~ \times 10^{22}~ cm^{-2}$. The 
estimate of $N_{H}^{~z}$ for source \#~1 is an indication of the 
presence of medium in the vicinity of this source (or inside the host 
galaxy) with a density comparable with the observed column density 
inside the disk and the bulge of our Galaxy.


\section{Conclusions}

 \begin{figure}[!ht]
   \centering \includegraphics[width=9cm,height=6cm]{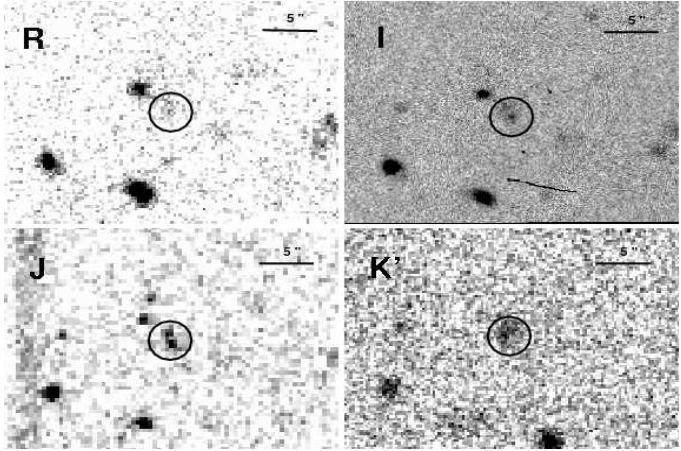}
       \caption{RIJK'-band images of the Chandra source \#~1. The
       field size is about $30^{\prime\prime}\times20^{\prime\prime}$,
       North is at top and East to the left. The black circle is
       $2^{\prime\prime}$ of radius and it is centered at coordinates
       reported in Table \ref{chandra} for this source.}
         \label{host}
   \end{figure}

We have performed a multiwavelength analysis of all the afterglow
candidates of GRB 031220 inside the HETE error circle. A deep
inspection of optical and infrared images taken at different epochs
did not reveal any new variable sources without X-ray emission. In the
optical vs. X-ray diagram our best afterglow candidate is located in
the region of dark events and show redder $R - K'$ colour than typical
optical transients. For these reasons we can infer that this GRB
belongs to the class of {\it Dark Bursts}.

If Chandra source \#~1 is the host galaxy of GRB 031220 it is evident
that the optical-infrared flux extinction observed could not be
ascribed to high redshift, because at $Z_{phot}$ $\sim 1.9$ the Lyman
break is at $\sim 2650$~$\AA$. The host galaxy of this burst
(Fig.\ref{host}) shows a red colour ($R - K'$ $\sim$ 5) and should be
classified as an Extreme Red Object (ERO). The infered rest-frame
column density $N_{H}^{~z}$ derived for this source permit us to
conclude that the {\it darkness} of GRB 031220 is the result of dust
extinction in the circum burst medium or inside the host galaxy.

\begin{acknowledgements} 

This study is partially supported by the Italian Space Agency (ASI)
the through the grant accorded to A.M. We thank H. Tananbaum and the
Chandra planning team for the successful implementation of the
follow-up observation. This study is partially based on data taken at
the 2.2m and 3.5m telescopes of the Centro Astron\'omico Hispanico
Alem\'an de Calar Alto, operated by the Max Plank Institute of
Hidelberg and Centro Superior de Investigaciones Cient\'{\i}ficas and
partially supported by the Spanish Ministry of Science and Education
through programmes ESP2002-04124-C03-01 and AYA2004-01515 (including
FEDER funds). This research is also partially supported by the EU FP5
RTN ``Gamma ray bursts: an enigma and a tool'', through the grant
accorded to B.G.

\end{acknowledgements}


\begin{thebibliography}{}

\bibitem[2000]{ander} Andersen M. I. {\it et al.}, 2000, A\&A 364, L54

\bibitem[2004]{anto} Antonelli L. A. {\it et al.}, 2003, GCN 2503

\bibitem[2003]{atteia} Atteia J. L., 2003, A\&A 407, L1

\bibitem[2002]{bagoly} Bagoly Z. {\it et al.}, 2003, A\&A 398, 919

\bibitem[2003]{bal03} Ballet J., 2003, C. Motch \& J.-M. Hameury eds,
      EAS Publications Series 7, 125

\bibitem[2002]{berger} Berger E. {\it et al.}, 2002, ApJ 581, 981 

\bibitem[1996]{bertin} Bertin E. \& Arnouts S., 1996, A\&AS 144, 363

\bibitem[1997]{cos97} Costa, E., {\it et al.}, 1997, Nature, 387, 783

\bibitem[2000]{csabai} Csabai I. {\it et al.}, 2000, AJ 119, 69

\bibitem[2002]{depasquale2} De Pasquale M. {\it et al.}, 2003, ApJ
       592, 1018

\bibitem[2003]{depasquale} De Pasquale M. {\it et al.}, 2003, GCN 2502

\bibitem[2005]{dep05} De Pasquale M., {\it et al.}, 2005, submitted to A\&A, astro-ph/0507708

\bibitem[1990]{dickey} Dickey J. M. \& Lockman F. J., 1990, ARA\&A,
       28, 215

\bibitem[2001]{djor} Djorgovski S. G. {\it et al.}, 2001, ApJ 562, 654

\bibitem[1999]{soto} Fernandez-Soto A. {\it et al.}, 1999, ApJ 513, 34

\bibitem[1997]{fioc} Fioc M. \& Rocca-Volmerange B., 1997, A\&A 326, 950

\bibitem[2001]{fynbo} Fynbo J. U. {\it et al.}, 2001, A\&A 369, 373

\bibitem[2000]{fontana} Fontana A. {\it et al.}, 2000, AJ 120, Issue
      5, 2206

\bibitem[2001]{galama} Galama T. \& Wijers R. A. M. J., 2001, ApJ 549, L209

\bibitem[2004]{gendre} Gendre B. {\it et al.}, 2004, GCN 2523

\bibitem[2000]{gendre2} Gendre B. \& Boer M, 2005, A\&A  430, 465

\bibitem[2005]{gen05} Gendre B., Corsi, A., Piro, L., 2005, submitted to A\&A

\bibitem[2004]{goro} Gorosabel J. {\it et al.}, 2004, GCN 2513

\bibitem[2003]{goro2} Gorosabel J. {\it et al.}, 2003, A\&A 400, 127

\bibitem[2002]{goro3} Gorosabel J. {\it et al.}, 2002, A\&A 384, 11

\bibitem[2003]{kosugi} Kosugi G. {\it et al.}, 2003, GCN 2497 

\bibitem[2004]{jako} Jakobsson P. {\it et al.}, 2004, ApJ 617, L21

\bibitem[2000]{lamb} Lamb D. Q. \& Reichart D. E., 2000, ApJ 536, 1

\bibitem[2002]{lazza} Lazzati D. {\it et al.}, 2002, MNRAS 330, 583

\bibitem[2002]{leborgne} Le Borgne D. \& Rocca-Volmerange B., 2002,
       A\&A 386, 446

\bibitem[2003]{lefloch} Le Floc'h E. {\it et al.}, 2003, A\&A 400, 499

\bibitem[1999]{macfad} MacFadyen A. I. \& Woosley S. E., 1999, ApJ 524, 262

\bibitem[2001a]{maio} Maiolino R. {\it et al.}, 2001a, A\&A 365, 28

\bibitem[1998]{pacz} Paczy\'nski B., 1998, ApJ 494, L45

\bibitem[1998]{pir98} Piro, L., {\it et al.}, 1998, A\&A, 331, L41

\bibitem[2002]{piro} Piro L. {\it et al.}, 2002, ApJ 577, 680

\bibitem[2003]{price} Price P. A. \& Peterson B. A., 2003, GCN 1987

\bibitem[2002]{reichart} Reichart D. E. \& Price P. A., 2002, ApJ
       565, 174

\bibitem[2004]{rykoff} Rykoff E. S. {\it et al.}, 2004, GCN 2495

\bibitem[2003]{robinson} Rowan-Robinson M., 2003, MNRAS 345, 819

\bibitem[1998]{sari} Sari R., Piran T. \& Narayan R., 1998, ApJ 497, L17

\bibitem[2001]{schaefer} Schaefer B. E., Deng M. \& Band D .L., 2001,
      ApJ 563, L123

\bibitem[1998]{schlegel} Schlegel D. J. {\it et al.}, 1998, ApJ 500, 525

\bibitem[2003]{smith} Smith D. A. {\it et al.}, 2003, ApJ 596, L151

\bibitem[1987]{stetson} Stetson P. B., 1987, PASP, 99, 191

\bibitem[2004]{stratta} Stratta G. {\it et al.}, 2004, ApJ 608, 846

\bibitem[2000]{wheeler} Wheeler J. C. {\it et al.}, 2000, ApJ 537, 810

\bibitem[1998]{wijers} Wijers R. A. M. J., Bloom J. S., Bagla J. S. \&
      Natarajan P., 1998, MNRAS 294, L13


\end{thebibliography}
\end{document}